\documentclass[11pt]{article}
\usepackage{moriond,epsfig,subfigure,floatflt}
\usepackage[english]{babel}

\def\PLB{{\em Phys. Lett.}  B}
\def\PRL{\em Phys. Rev. Lett.}
\def\PRD{{\em Phys. Rev.} D}

\def\be{\begin{equation}}
\def\ee{\end{equation}}
\def\bea{\begin{eqnarray}}
\def\eea{\end{eqnarray}}

\begin{document}
\vspace*{4cm}
\title{THE COMPASS SPIN PHYSICS PROGRAM}

\author{ S. PANEBIANCO, on behalf of the COMPASS collaboration }

\address{CEA-Saclay, DSM/DAPNIA/SPhN, 91191 Gif/Yvette, France}

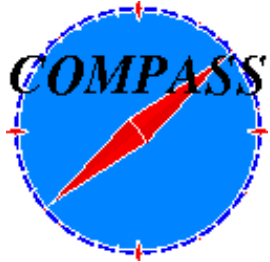
\maketitle\abstracts{
One of the main aims of the COMPASS experiment at CERN is the study of the spin structure of the nucleon and in particular the determination of the gluon polarization in the nucleon. We present some new results of 2002-2003 data analysis. They concern a precise measurement of the deuteron structure function $g_1$ at small $x$, some preliminary result on Collins and Sivers asymmetries which are linked to transversity and a new measurement of $\Delta G/G$. 
}

\section{General overview}

The structure of the nucleon in terms of its constituents is described by the \emph{Parton Distribution Functions} (PDF). They are used to describe the soft part of the deep inelastic scattering (DIS) process, where a lepton scatters on a nucleon with the exchange of a high virtual photon. In unpolarized DIS one has access to the quark and gluon PDF, $q(x,Q^2)$ and $G(x,Q^2)$. On the contrary, in polarized DIS two other PDF can be measured: $\Delta q=q^{\uparrow\uparrow}-q^{\uparrow\downarrow}$ and $\Delta G=G^{\uparrow\uparrow}-G^{\uparrow\downarrow}$. The polarized PDF are the fundamental pieces to understand the nucleon spin. For a nucleon with a helicity $+1/2$ we have: $\frac{1}{2}=\frac{1}{2}\Delta\Sigma+\Delta G + L_z^q + L_z^g$, where $\frac{1}{2}\Delta\Sigma=\frac{1}{2}(\Delta u+\Delta d+\Delta s)$ is the contribution from the spin of quarks, $\Delta G$ is the contribution from the gluon spin and $L_z^q + L_z^g$ is the total orbital angular momentum of quarks and gluons. The theoretical prediction for $\Delta\Sigma$, coming from the parton model, is $\Delta\Sigma\sim0.6$, whereas EMC, SMC and SLAC measured $\Delta\Sigma=0.27\pm0.13$ \cite{SMC}. It is a surprising result suggesting that the spin of the quarks accounts for only a small fraction of the nucleon spin. A measurement of $\Delta G$ is therefore mandatory. Both $\Delta\Sigma$ and $\Delta G$ are accessible in polarized DIS experiments by measuring respectively inclusive and semi-inclusive double spin asymmetries.

COMPASS \cite{PROP} is a polarized deep inelastic scattering experiment installed at CERN SPS and uses a 160~GeV longitudinally polarized muon beam with an intensity of $2\times 10^8\mu^+/$spill. The polarized deuteron ($^6$LiD) target consists of an upstream and a downstream cell with opposite polarization. The particles produced in the interaction are detected behind the target in a two-stage spectrometer with high momentum resolution and high rate capability. COMPASS took data from 2002 to 2004, accumulating an integrated luminosity of $\sim 4.6$~fb$^{-1}$. We present here some new results related to the nucleon spin structure, mostly based on 2002-2003 statistics, corresponding to $\sim 2.2$~fb$^{-1}$.

\section{Virtual photon-deuteron asymmetry $A_1^d$} \label{sec:A1}

Information about helicity quark PDF can be extracted from the inclusive asymmetry $A_1^d$ which is derived from the cross section asymmetry $A^{\mu d}_{LL}$ through $DA_1^d\simeq A^{\mu d}_{LL}$:
\begin{equation}
A_1^d=\frac{\sum_q e_q^2(\Delta q + \Delta\bar{q})}{\sum_q e_q^2(q + \bar{q})}=\frac{1}{D}\frac{1}{P_bP_tf}\frac{1}{2}\left( \frac{N_u^{\uparrow\uparrow}-N_d^{\uparrow\downarrow}}{N_u^{\uparrow\uparrow}+N_d^{\uparrow\downarrow}}+\frac{N_d^{\uparrow\uparrow}-N_u^{\uparrow\downarrow}}{N_d^{\uparrow\uparrow}+N_u^{\uparrow\downarrow}} \right) ,
\end{equation}
\begin{floatingfigure}[r]{0.5\textwidth}
\begin{center}
\epsfig{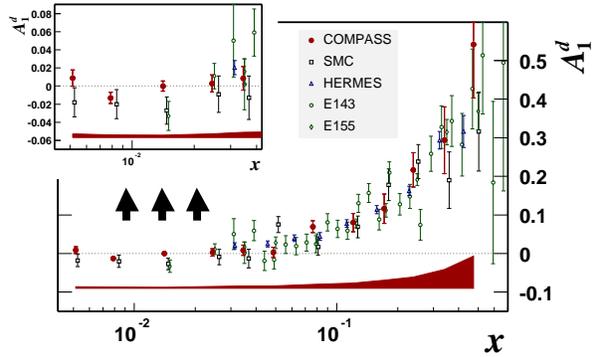}
\caption{The $A_1^d(x)$ asymmetry measured in COMPASS and previous results from SMC, HERMES and SLAC.}
\label{fig:A1}
\end{center}
\end{floatingfigure}
\noindent where N is the counting rate ($u$ and $d$ denote the two target cells, $\uparrow\uparrow$ and $\uparrow\downarrow$ denote the two longitudinal spin configurations of the target), $D$ is the depolarization factor from the muon to the virtual photon, $P_b$ is the beam polarization ($\sim 76\%$), $P_t$ and $f$ are the $^6$LiD target polarisation ($\sim 50\%$) and the process dependent dilution factor ($\sim 40\%$). 
The data selection is the standard inclusive DIS, with $Q^2>1$~GeV$^2$. The results from 2002 and 2003 data \cite{A1COMP} are shown in Fig. \ref{fig:A1}, together with previous results from SMC \cite{SMC}, HERMES \cite{HERMES} and SLAC \cite{SLAC}. Owing to the high luminosity and a high dilution factor for a solid state target, they significantly improve the accuracy in the region $x<0.03$. The achieved accuracy very much constrains the $g_1$ structure function extrapolation leading to a better precision in the determination of $\Delta \Sigma$.

\section{The $\Delta G/G$ measurement} \label{sec:DeltaG}

\begin{figure}[h]
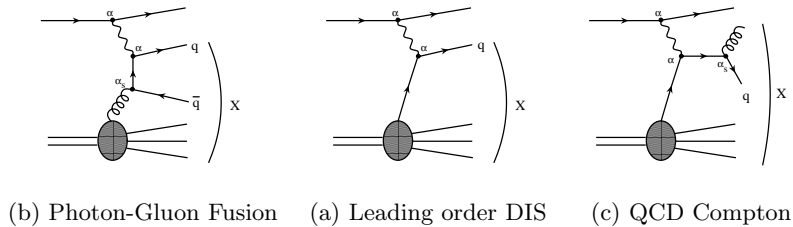

\begin{center}
\begin{tabular}{ccc}
\epsfig{file=eps/pgf_1.eps,width=3cm,bbllx=2,bblly=605,bburx=232,bbury=824}&
\epsfig{file=eps/lodis_1.eps,width=3cm,bbllx=2,bblly=605,bburx=232,bbury=824}&
\epsfig{file=eps/qcdc_1.eps,width=3cm,bbllx=2,bblly=605,bburx=232,bbury=824}\\
\footnotesize{(b) Photon-Gluon Fusion}   &\footnotesize{(a) Leading order DIS}   &\footnotesize{(c) QCD Compton}
\end{tabular}
\caption{At the order $\alpha_s$, three processes contribute to the cross section for lepton-nucleon scattering.}
\label{fig:proc}
\end{center}
\end{figure}

In COMPASS, we access the gluon polarization via the photon-gluon fusion (PGF) process, whereby a virtual photon couples to a gluon via a quark, leading to the production of a $q\bar{q}$ pair. From the counting rate asymmetry between two target cells, one extracts the helicity asymmetry $A_{LL}^{\mu d}$ of the muon-deuteron cross section. If the muon-deuteron scattering involves a hard probe, the factorization theorem implies that the muon interacts perturbatively with a parton in the nucleon. At tree level, the three direct processes illustrated in figure \ref{fig:proc} contribute to $A_{LL}^{\mu d}$. 
The gluon polarization in the nucleon, $\Delta G/G$, is extracted from the cross section asymmetry of the photon-gluon fusion: $A_{LL}^{\mu d(pgf)}=\langle\hat{a}_{LL}^{pgf}\rangle\Delta G/G$, where the asymmetry $\langle\hat{a}_{LL}^{pgf}\rangle$ of the partonic (photon-gluon) reaction can be calculated. It is of course essential to get rid of the dominant background processes, such as the leading order DIS (LODIS) and the QCD Compton (QCDC). In COMPASS this is done in two ways, through the selection of high transverse momentum hadron pairs or of open charm events.

\subsection{High $p_t$ hadron pair leptoproduction} \label{subsec:highpt}

The selection of events with a pair of high $p_t$ hadrons removes most of the LODIS process, which is the dominant source of background to the PGF in the $Q^2>1$~GeV$^2$ region. It requires the $p_t$ of the two leading hadrons to be larger than $0.7$~GeV and $p_{t1}^2+p_{t2}^2>2.5$~GeV$^2$. However, the QCD Compton process remains, and a Monte-Carlo simulation is necessary to estimate its contribution to the asymmetry. The helicity asymmetry in the production of high $p_t$ hadron pairs from 2002 and 2003 data at $Q^2>1$~GeV$^2$ is $A_{LL}^{\gamma^{\star}d}=-0.015\pm0.080_{stat.}\pm0.013_{syst.}$. The systematic error takes into account false asymmetries due to spectrometer and target effects. The gluon polarization is then determined from the asymmetry, assuming LODIS and QCDC contributions to be negligible and to be taken into account in the systematic error: $\frac{\Delta G}{G}=0.06\pm0.31_{stat.}\pm0.06_{syst.}$ at $x_g=0.13\pm0.08$(RMS), where $x_g$ is the fraction of nucleon momentum carried by gluons.
\begin{floatingfigure}[r]{0.5\textwidth}
\begin{center}
\epsfig{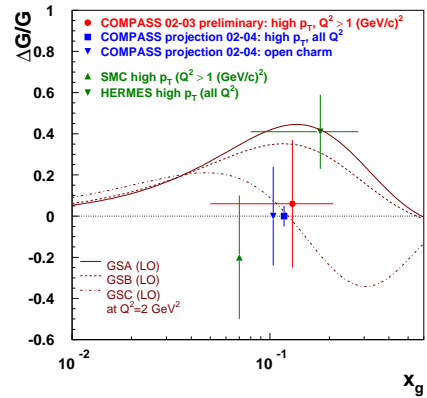}
\caption{The gluon polarization measured in COMPASS and previous results from SMC and HERMES.}
\label{fig:DG}
\end{center}
\end{floatingfigure}
\noindent Fig. \ref{fig:DG} shows the COMPASS result compared with previous HERMES \cite{HERMES2} and SMC \cite{SMC2} results. The discrepancy with HERMES measurement arises from the fact that  HERMES explored also the $Q^2<1$~GeV$^2$ region where the additional contributions from resolved processes, in which the virtual photon fluctuates to hadronic state (VMD or $q\bar{q}$), have been neglected. COMPASS showed indeed that these contributions cannot be neglected. The analysis of the 2002-2003 data in the $Q^2<1$~GeV$^2$ range is almost finished, leading to a reduction of the statistical uncertainty by at least a factor of four.

\subsection{Open charm leptoproduction} \label{subsec:charm}

Due to its large mass ($\sim 1.5$~GeV), the $c$ quark has a very low probability to be present in the nucleon's sea or to be produced in the fragmentation. Moreover, at tree level, $c$ quarks can only be produced in the photon-gluon fusion. Therefore, detecting a charmed hadron in the final state is a very clean way to select PGF events. This is done by reconstructing $D^0$ and $D^{\star}$ mesons from their hadronic decay products: $D^0\rightarrow K^- \pi^+$ and $D^{\star}\rightarrow D^0 \pi^+_s \rightarrow K^- \pi^+ \pi^+_s$ (and the charge conjugated processes). The difference between $D^{\star}$ and $D^0$ masses is only 145~MeV, thus the available phase space for the soft pion $\pi^+_s$ is small in this decay. Therefore the combinatorial background is strongly suppressed. Additional cuts on the momentum fraction of the D-meson, on the decay angle distribution and on the kaon and pion identification help to further reduce the background. 
Table \ref{tab:charm} presents the counting rate asymmetry of untagged $D^0$ and of $D^{\star}$ tagged events, obtained from 2002 and 2003 data. Despite the large number of $D^0$, leading to a smaller statistical error on the raw asymmetry, the so called effective signal $S_{eff}=\frac{S}{1+B/S}$, which is proportional to inversed squared statistical accuracy on $\Delta G/G$, is higher in the $D^{\star}$ case because of the better $S/B$ ratio. The extraction of $\Delta G/G$ is challenging because of the limited statistics. The analysis is ongoing and the expected accuracy from 2002, 2003 and 2004 data is $0.24$.
\begin{table}[h]
\begin{center}
\caption{Counting rate asymmetry of untagged $D^0$ and of $D^{\star}$ tagged events and corresponding $S_{eff}$.\label{tab:charm}}
\vspace{0.4cm}
\begin{tabular}{|c|c|c|}
\hline
Channel &Raw asymmetry  & $S_{eff}$\\
\hline
\hline
Untagged $D^0$  &$0.001 \pm 0.004$ & $391 \pm 26 $ \\
\hline
$D^{\star}$ tagged $D^0$  &$-0.04 \pm 0.02$ & $704 \pm 69 $ \\
\hline
\end{tabular}
\end{center}
\end{table}

\subsection{Looking for transversity} \label{subsec:transv}

\begin{figure}[h]
\begin{center}
\epsfig{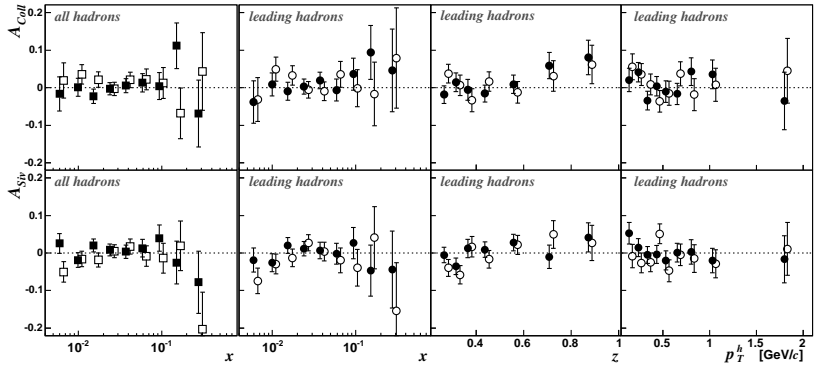}
\caption{Collins (top) and Sivers (bottom) asymmetries against $x$, $z$ and $p_T^h$ for positive (full points) and negative (open points) hadrons. The first column gives the asymmetries for all hadrons, the other three columns for leading hadrons. Error bars are statistical only.}
\label{fig:transv}
\end{center}
\end{figure}

The third fundamental PDF of the quark is the transversity distribution $\Delta_T q(x)$. Because of its odd chirality, this distribution does not contribute to inclusive DIS. It may instead be extracted from measurements of the spin asymmetries in cross sections for semi inclusive DIS of a lepton on transversely polarized nucleon. The $\Delta_T q(x)$ PDF can be measured in combination with a chiral-odd fragmentation function $D_{Tq}^h(z,\vec{p}_T^h)$, via azimuthal single spin asymmetries in the hadronic end-product (Collins effect). A similar effect can arise from a possible $k_T$  structure of a transversely polarized nucleon (the Sivers function $\Delta_0^T q$), which also causes an azimuthal asymmetry in the produced hadrons. Lepton scattering on transversely polarized nucleons is a favorable setting to disentangle the Collins and Sivers effects since they show a dependence from linearly independent kinematics variables. The results of the azimuthal asymmetries from hadron muon-production on a transversely polarized deuteron target, plotted against the kinematic variables $x$, $z$ and $p_T^h$ are shown in Fig. \ref{fig:transv} (2002 data only). Within the statistical accuracy of the data, both the Collins and Sivers asymmetries turned out to be small and compatible with zero \cite{TRANSV}.


\end{document}